\documentclass{article}

\usepackage{ifthen,amsmath,amssymb,latexsym,epsf,stmaryrd}
\newboolean{icmpstyle}
\newboolean{draft}

\setboolean{icmpstyle}{true}
\setboolean{draft}{false}

\ifthenelse{\boolean{draft}}{\usepackage{showkeys}}{}

\ifthenelse{\boolean{icmpstyle}}{ 
 } 
{                                      
\usepackage{amsthm,amscd,times}
} 

\def\oline{                          
\parbox{3mm}{\begin{picture}(5,10)   
\put(1,8){$\bullet$}                 
\put(1,1){$\bullet$} \put(2,9){\line(0,-1){7}}
\end{picture}}}
\def\oc{                     
\parbox{10mm}{\begin{picture}(10,10) 
\put(4.5,8){$\bullet$}               
\put(1,1){$\bullet$} \put(8,1){$\bullet$}
\put(5.5,9){\line(1,-2){3.5}} \put(5.5,9){\line(-1,-2){3.5}}
\end{picture}}}
\def\ofork{                           
\parbox{10mm}{\begin{picture}(10,17)  
\put(4.5,15){$\bullet$}               
\put(4.5,8){$\bullet$}                
\put(8,1){$\bullet$}                  
\put(1,1){$\bullet$} \put(5.5,9){\line(0,1){7}}
\put(5.5,9){\line(-1,-2){3.5}} \put(5.5,9){\line(1,-2){3.5}}
\end{picture}}}
\def\olongline{                       
\parbox{3mm}{\begin{picture}(3,17)    
\put(1,15){$\bullet$}                 
\put(1,8){$\bullet$}                  
\put(1,1){$\bullet$}                  
\put(2,2){\line(0,1){14}}
\end{picture}}}
\def\ocf{                                     
\parbox{16mm}{\begin{picture}(16,15)
\put(8,12){$\bullet$} 
\put(9,13){\line(1,-1){7}} \put(9,13){\line(-1,-1){7}}
\put(9,13){\line(1,-2){3.5}} \put(9,13){\line(0,-1){7}}
\put(9,13){\line(-1,-2){3.5}} \put(1,2){$\tau_1$}
\put(7,2){$\ldots$} \put(15,2){$\tau_n$}
\end{picture}}}

\def\wheel{\;\raisebox{-0.6cm}{\epsfysize=1.5cm\epsfbox{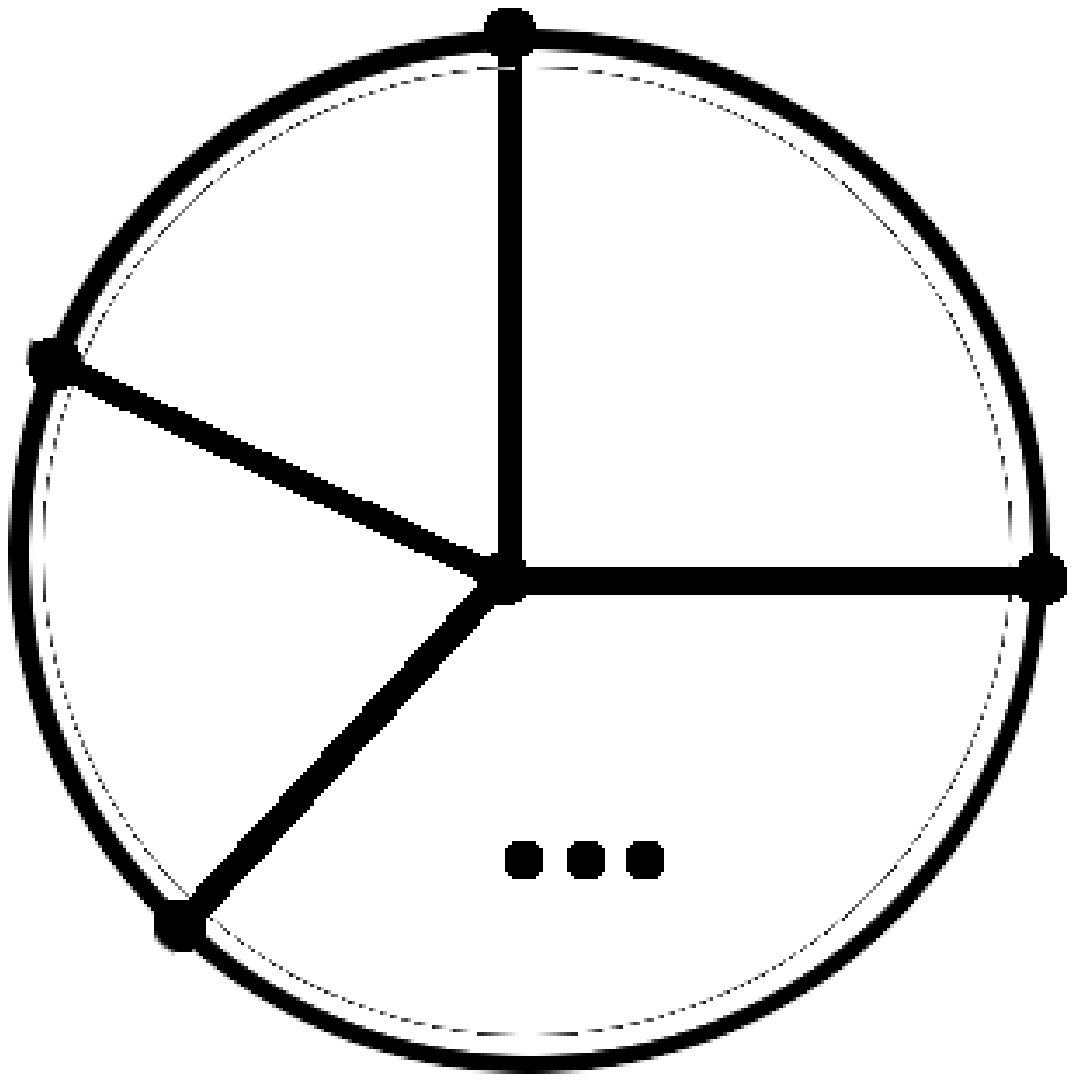}}\;}
\def\seone{\;\raisebox{-0.22cm}{\epsfysize=0.6cm\epsfbox{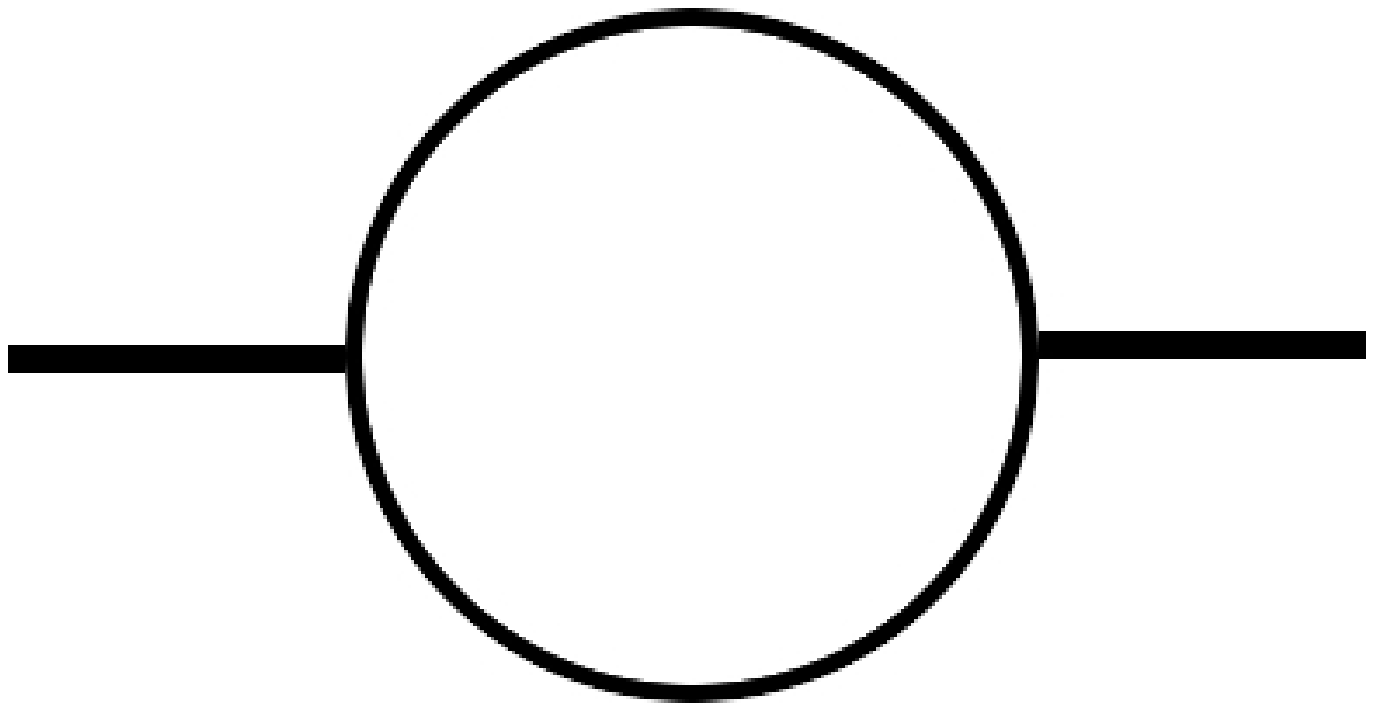}}\;}
\def\setwo{\;\raisebox{-0.22cm}{\epsfysize=0.6cm\epsfbox{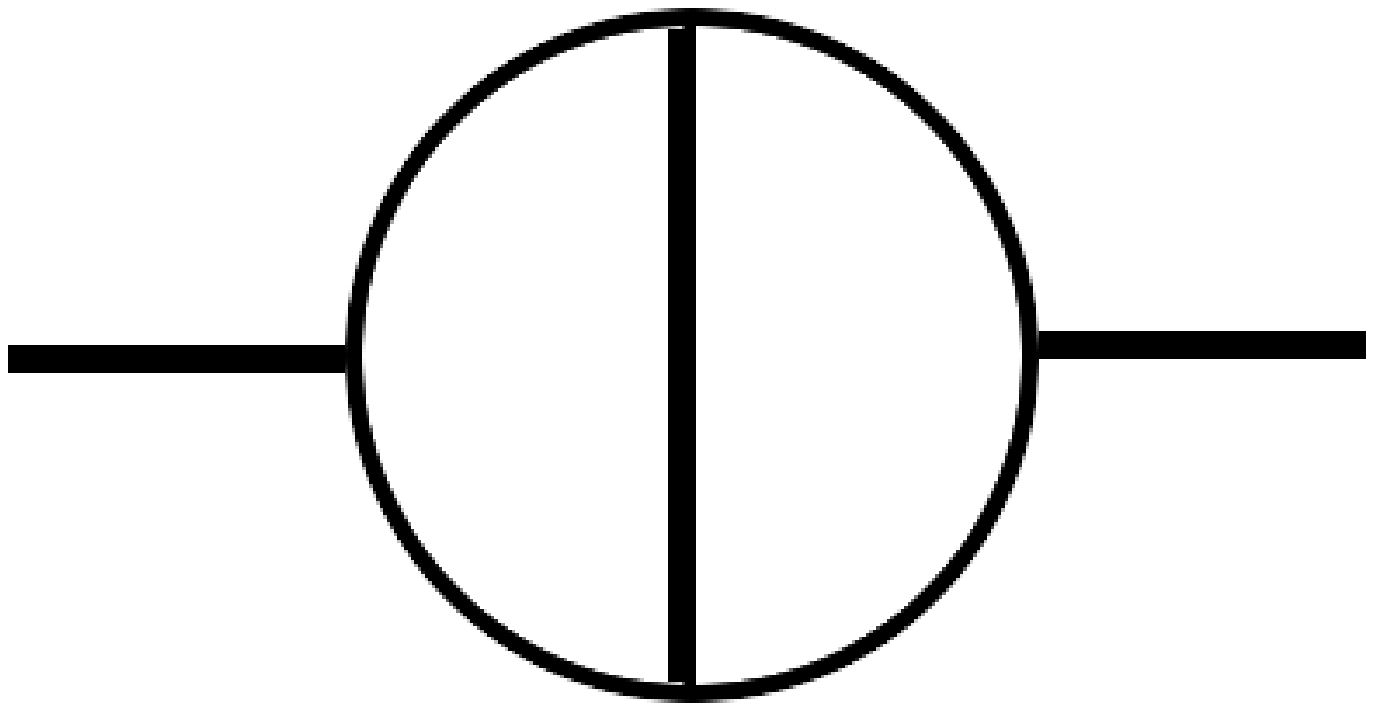}}\;}
\def\vert{\;\raisebox{-0.4cm}{\epsfysize=1cm\epsfbox{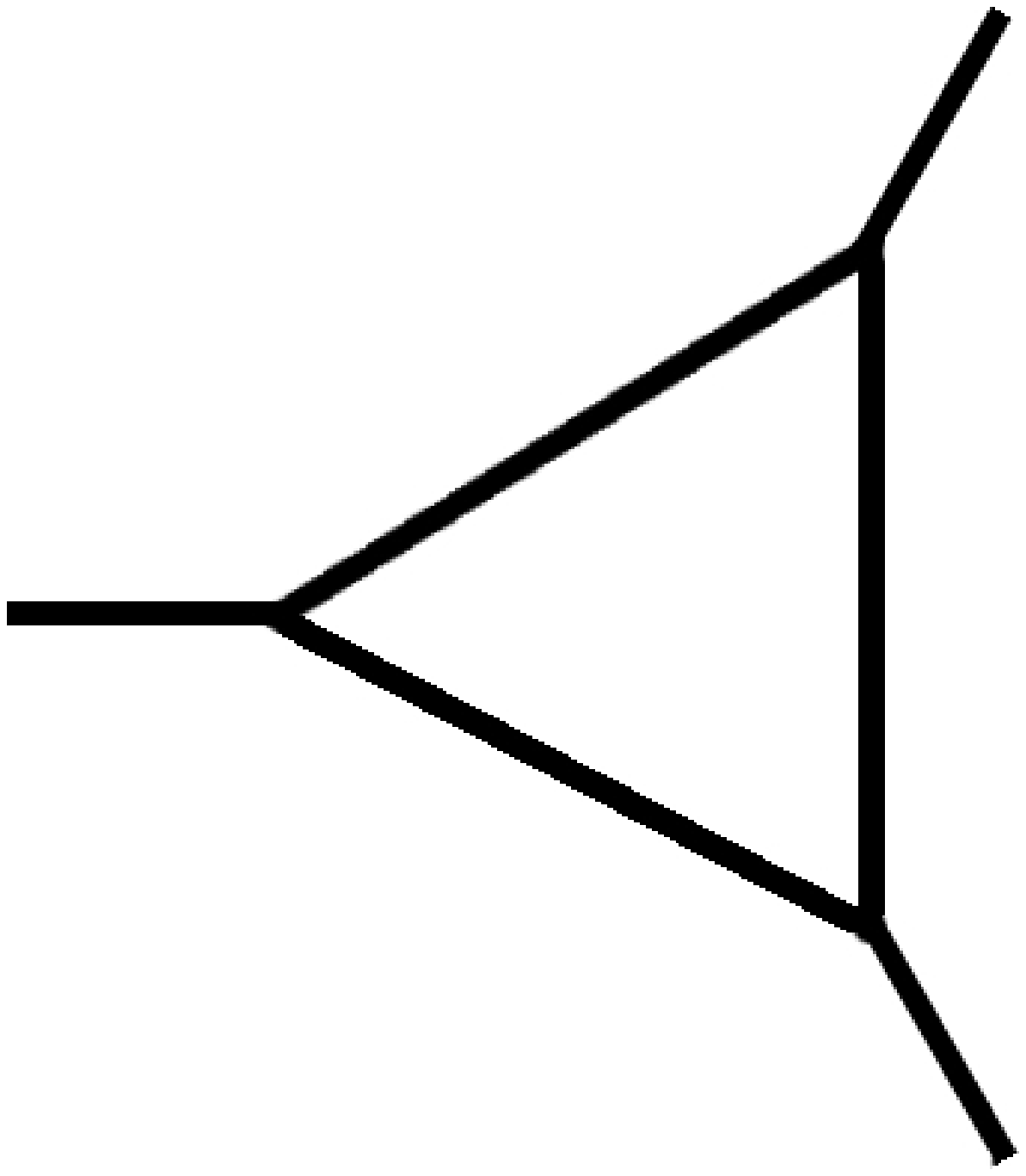}}\;}
\def\vertiu{\;\raisebox{-0.4cm}{\epsfysize=1cm\epsfbox{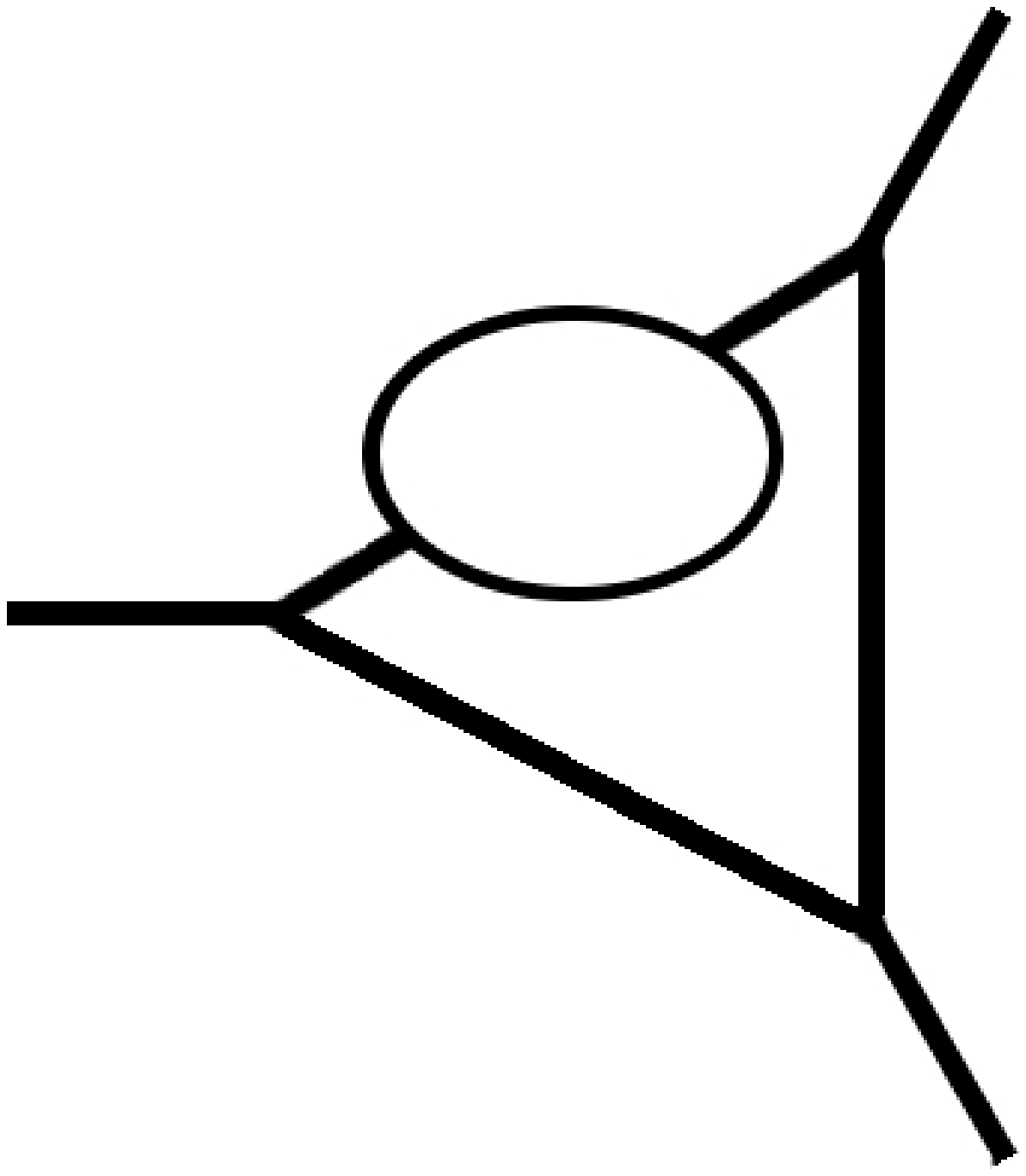}}\;}
\def\vertid{\;\raisebox{-0.4cm}{\epsfysize=1cm\epsfbox{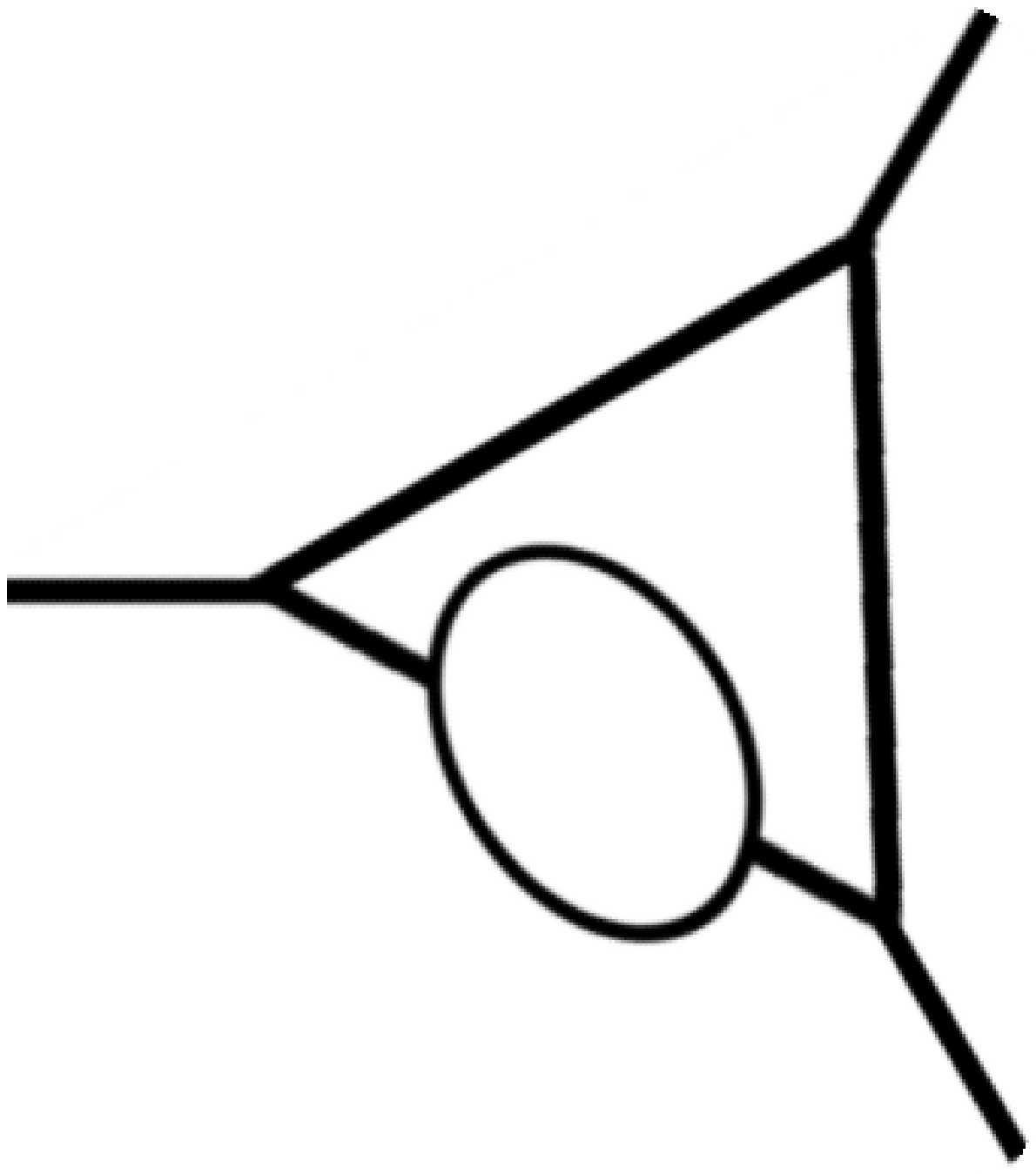}}\;}
\def\vertir{\;\raisebox{-0.4cm}{\epsfysize=1cm\epsfbox{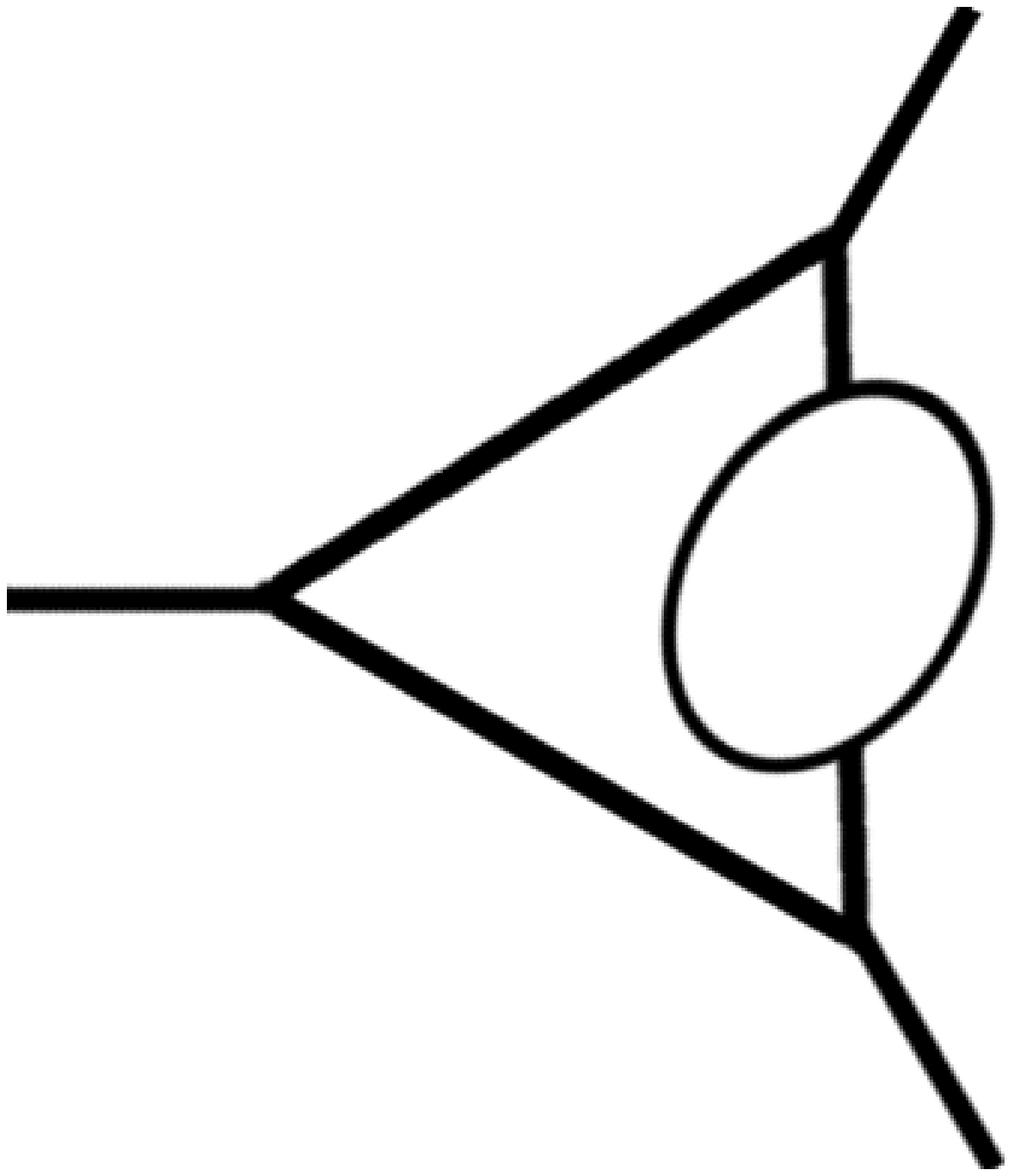}}\;}
\def\phifull{\;\raisebox{-0.65cm}{\epsfysize=1.5cm\epsfbox{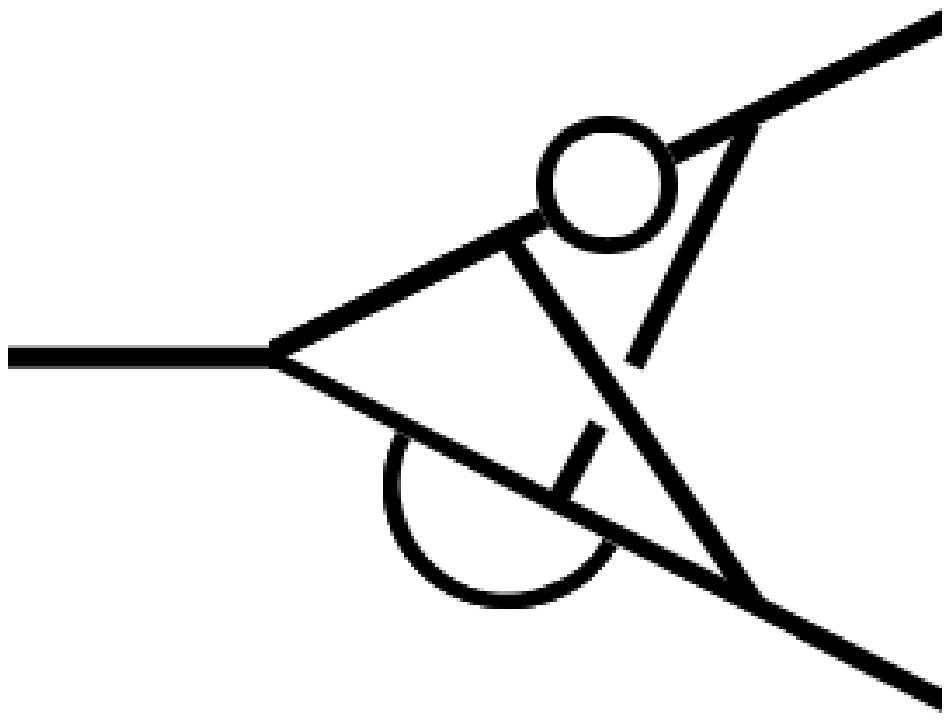}}\;}
\def\phitop{\;\raisebox{-0.65cm}{\epsfysize=1.5cm\epsfbox{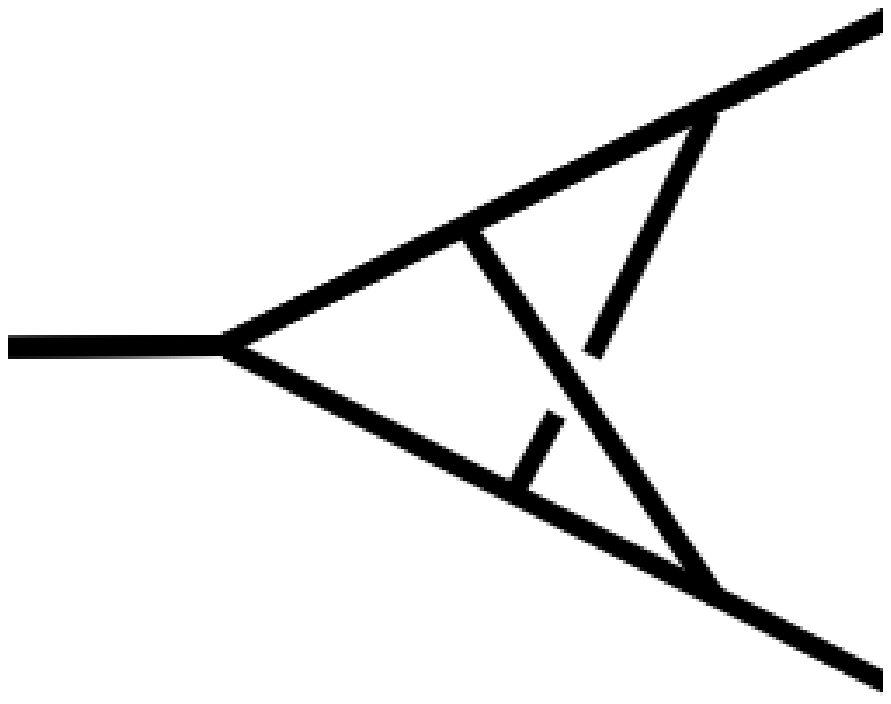}}\;}
\begin{document}
\ifthenelse{\boolean{icmpstyle}}{ 
\title{New algebraic aspects of perturbative and non-perturbative Quantum Field Theory}
\author{Christoph Bergbauer$\left.^{1,4}\right.$ and Dirk Kreimer$\left.^{2,3}\right.$\\[7mm]
$\left.^1\right.$Freie Universit\"at Berlin, Institut f\"ur Mathematik II\\
Arnimallee 3, 14195 Berlin, Germany\\[2mm]
$\left.^2\right.$CNRS at Institut des Hautes Etudes Scientifiques\\
35 route de Chartres, 91440 Bures-sur-Yvette, France\\[2mm]
$\left.^3\right.$Boston University, Center for Mathematical Physics\\
111 Cummington Street, Boston, MA 02215, USA\\[2mm]
$\left.^4\right.$Erwin-Schr\"odinger-Institut\\
Boltzmanngasse 9, 1090 Wien, Austria\\[2mm]
\texttt{bergbau@math.fu-berlin.de, kreimer@ihes.fr}}
\date{April 1, 2007}
\maketitle
\begin{abstract}
In this expository article we review recent advances in our
understanding of the combinatorial and algebraic structure of
perturbation theory in terms of Feynman graphs, and Dyson-Schwinger
equations. Starting from Lie and Hopf algebras of Feynman graphs,
perturbative renormalization is rephrased algebraically. The
Hochschild cohomology of these Hopf algebras leads the way to
Slavnov-Taylor identities and Dyson-Schwinger equations. We discuss
recent progress in solving simple Dyson-Schwinger equations in the
high energy sector using the algebraic machinery. Finally there
is a short account on a relation to algebraic geometry and number
theory: understanding Feynman integrals as periods of mixed (Tate) motives.
\end{abstract}
}
{ 
\title[New algebraic aspects of perturbative and non-perturbative
QFT]{New algebraic aspects of perturbative and non-perturbative
Quantum Field Theory}
\author{Christoph Bergbauer and Dirk Kreimer}
\address{Freie Universit\"at Berlin\\Institut f\"ur Mathematik II\\ Arnimallee 3\\ 14195
Berlin\\ Germany and
\newline
Erwin-Schr\"odinger-Institut\\Boltzmanngasse 9\\1090 Wien\\Austria}
\email{bergbau@math.fu-berlin.de}
\address{CNRS at Institut des Hautes \'Etudes Scientifiques\\35 route de Chartres\\91440 Bures-sur-Yvette\\France and
\newline
Center for Mathematical Physics\\Boston University\\111 Cummington
Street\\Boston, MA 02215\\USA} \email{kreimer@ihes.fr}
\ifthenelse{\boolean{draft}}{\date{(Draft) \today}}{\date{April 1, 2007}}
\begin{abstract}
In this expository article we review recent advances in our
understanding of the combinatorial and algebraic structure of
perturbation theory in terms of Feynman graphs, and Dyson-Schwinger
equations. Starting from Lie and Hopf algebras of Feynman graphs,
perturbative renormalization is rephrased algebraically. The
Hochschild cohomology of these Hopf algebras leads the way to
Slavnov-Taylor identities and Dyson-Schwinger equations. We discuss
recent progress in solving simple Dyson-Schwinger equations in the
high energy sector using the algebraic machinery. Finally there
is a short account on a relation to algebraic geometry and number
theory: understanding Feynman integrals as periods of mixed (Tate) motives.
\end{abstract}
\maketitle
\setcounter{tocdepth}{2} \tableofcontents } 
\section{Introduction}
As \ifthenelse{\boolean{icmpstyle}}{} 
{
\footnote{Contribution to the Proceedings of the
International Congress on Mathematical Physics 2006, Rio de Janeiro.
Based on a talk given by the first named author in the Quantum Field
Theory session.}}
elements of perturbative expansions of Quantum field theories,
Feynman graphs have been playing and still play a key role both for our
conceptual understanding and for state-of-the-art computations in particle
physics. This article is concerned with several aspects of Feynman graphs:
First, the combinatorics of perturbative renormalization give rise to
Hopf algebras of rooted trees and Feynman graphs. These Hopf algebras
come with a cohomology theory and structure maps that
help understand important physical notions, such as
locality of counterterms, the beta function, certain symmetries, or
Dyson-Schwinger equations from a unified mathematical point of view.
This point of view is about self-similarity and recursion.
The atomic (primitive) elements in this combinatorial approach are
divergent graphs without subdivergences. They must be studied by
additional means, be it analytic methods or algebraic geometry and number
theory, and this is a significantly more difficult task. However,
the Hopf algebra structure of graphs for renormalization is in this sense
a substructure of the Hopf algebra structure underlying the relative
cohomology of graph hypersurfaces needed to understand the number-theoretic
properties of field theory amplitudes \cite{BlochJapan,BlochToronto}.
\section{Lie and Hopf algebras of Feynman graphs}
Given a Feynman graph $\Gamma$ with several divergent subgraphs, the
Bogoliubov recursion and Zimmermann's forest formula tell how $\Gamma$
must be renormalized in order to obtain a finite conceptual result,
using only local counterterms. This has an analytic
(regularization/extension of distributions) and a combinatorial aspect.
The basic \emph{combinatorial} question of perturbative
renormalization is to find a good model which describes
disentanglement of graphs into subdivergent pieces, or dually insertion
of divergent pieces one into each other, from the point of view of
renormalized Feynman rules. It has been known now for several years that
commutative Hopf algebras and (dual) Lie algebras provide such a framework
\cite{Kreimer,CK,CK2} with many ramifications in pure mathematics.
From the physical side, it is important to know that, for example, recovering
aspects of gauge/BRST symmetry \cite{VP,Suijlekom1,Anatomy,Suijlekom2} and the
transition to nonperturbative equations of motion
\cite{BroadhurstKreimerDSE,Kreimer2,KreimerTrouble,MK2,BK2,KY,KreimerToronto,KY2,BKW}
are conveniently possible in this framework, as will be discussed in
subsequent sections. \\\\
In order to introduce these Lie and Hopf algebras, let us now fix a renormalizable
quantum field theory (in the sense of perturbation theory), given by a local
Lagrangian. A convenient first example is massless $\phi^3$ theory in 6 dimensions.
We look at its perturbative expansion in terms of 1PI Feynman graphs. Each 1PI graph
$\Gamma$ comes with two integers, $|\Gamma|=\operatorname{rank} H_1(\Gamma),$ its number of loops, and
$\operatorname{sdd}(\Gamma),$ its superficial degree of divergence. As usual,
vacuum and tadpole graphs need not be considered, and the only remaining superficial
divergent graphs have exactly two or three external edges, a feature of
renormalizability. Graphs without subdivergences are called \emph{primitive}.
Here are two examples.
\begin{equation*}
\phifull\quad\phitop
\end{equation*}
Both are superficially divergent as they have three external edges.
The first one has two subdivergences, the second one is primitive.
Note that there are infinitely many primitive graphs with three external edges.
In particular, for every $n\in\mathbb{N}$ one finds a primitive $\Gamma$ such that $|\Gamma|=n.$ \\\\
Let now $L$ be the $\mathbb{Q}$-vector space generated by all the superficially divergent
$(\operatorname{sdd}\ge 0)$ 1PI graphs of our theory, graded by the number of loops
$|\cdot|. $ There is an operation on $L$ given by insertion of graphs into each other:
Let $\gamma_1,\gamma_2$ be two generators of $L.$ Then
\begin{equation*}
\gamma_1\star\gamma_2 := \sum_\Gamma n(\gamma_1,\gamma_2,\Gamma)
\end{equation*}
where $n(\gamma_1,\gamma_2,\Gamma)$ is the number of times that $\gamma_1$ shows up as a
subgraph of $\Gamma$ and $\Gamma/\gamma_1\cong \gamma_2.$ Here are two examples:
\begin{eqnarray*}
\seone\star\vert &=& \vertir+\vertiu+\vertid\\
\vert\star\seone &=& 2 \setwo
\end{eqnarray*}
This definition is extended bilinearly
onto all of $L.$ Note that $\star$ respects the grading as $|\gamma_1\star\gamma_2|=|\gamma_1|+|\gamma_2|.$
The operation $\star$ is not in general associative. Indeed, it is pre-Lie \cite{CK,CKInsElim}:
\begin{equation}\label{eq:preLie}
(\gamma_1\star\gamma_2)\star \gamma_3 - \gamma_1\star(\gamma_2\star\gamma_3) = (\gamma_1\star\gamma_3)\star\gamma_2 -
\gamma_1\star(\gamma_3\star\gamma_3).
\end{equation}
To see that (\ref{eq:preLie}) holds observe that on both sides nested
insertions cancel. What remains are disjoint insertions of $\gamma_2$
and $\gamma_3$ into $\gamma_1$ which do obviously not depend on the
order of $\gamma_2$ and $\gamma_3.$ One defines a Lie bracket on $L:$
\begin{equation*}
[\gamma_1,\gamma_2] := \gamma_1\star\gamma_2-\gamma_2\star\gamma_1.
\end{equation*}
The Jacobi identity for $[\cdot,\cdot]$ is satisfied as a consequence
of the pre-Lie property (\ref{eq:preLie}) of $\star.$ This makes $L$
a graded Lie algebra. The bracket is defined by mutual insertions of
graphs. As usual, $\mathcal{U}(L),$ the universal envelopping algebra
of $L$ is a cocommutative Hopf algebra. Its graded dual, in the sense
of Milnor-Moore, is therefore a commutative Hopf algebra $\mathcal{H}.$
As an algebra, $\mathcal{H}$ is free commutative, generated by the
vector space $L$ and an adjoined unit $\mathbb{I}.$ By duality, one expects the coproduct of $\mathcal{H}$
to disentangle its argument into subdivergent pieces. Indeed, one finds
\begin{equation}\label{eq:graphdelta}
\Delta(\Gamma) = \mathbb{I}\otimes \Gamma+\Gamma\otimes \mathbb{I} +
\sum_{\gamma\subsetneq \Gamma} \gamma\otimes \Gamma/\gamma.
\end{equation}
The relation $\gamma\subsetneq\Gamma$ refers to disjoint unions $\gamma$
of 1PI superficially divergent subgraphs of $\Gamma.$ Disjoint unions of
graphs are in turn identified with their product in $\mathcal{H}.$ For example,
\begin{equation*}
\Delta\left(\setwo\right) = \mathbb{I}\otimes\setwo+\setwo\otimes\mathbb{I}+2\vert\otimes\seone.
\end{equation*}
The coproduct respects the grading by the loop number, as does the
product (by definition). Therefore $\mathcal{H}=\bigoplus_{n=0}^\infty \mathcal{H}_n$
is a graded Hopf algebra. Since $\mathcal{H}_0\cong \mathbb{Q}$ it is connected.
The counit $\epsilon$ vanishes on the subspace $\bigoplus_{n=1}^\infty \mathcal{H}_n,$
called augmentation ideal, and $\epsilon(\mathbb{I})=1.$ As usual, if
$\Delta(x)=\mathbb{I}\otimes x+x\otimes \mathbb{I},$ the element $x$ is called \emph{primitive}.
The linear subspace of primitive elements is denoted $\operatorname{Prim}\mathcal{H}.$\\\\
The interest in $\mathcal{H}$ and $L$ arises from the fact
that the Bogoliubov recursion is essentially solved by the antipode
of $\mathcal{H}.$ In any connected graded bialgebra, the antipode $S$ is given
by
\begin{equation}\label{eq:antipode}
S(x) = -x - \sum S(x') x'', \quad x\notin \mathcal{H}_0
\end{equation}
in Sweedler's notation. Let now $V$ be a $\mathbb{C}$-algebra. The space
of linear maps $\mathcal{L}_\mathbb{Q}(\mathcal{H},V)$ is equipped with a
convolution product $(f,g)\mapsto f\ast g = m_V(f\otimes g)\Delta$ where
$m_V$ is the product in $V.$ Relevant examples for $V$ are suggested by regularization
schemes such as the algebra $V=\mathbb{C}[[\epsilon,\epsilon^{-1}]$ of Laurent series
with finite pole part for dimensional regularization (space-time dimension
$D=6+2\epsilon.$) The (unrenormalized) Feynman rules provide then an algebra
homomorphism $\phi: \mathcal{H}\rightarrow V$ mapping Feynman graphs to
Feynman integrals in $6+2\epsilon$ dimensions. On $V$ there is a linear
endomorphism $R$ (renormalization scheme) defined, for example minimal
subtraction $R(\epsilon^n)=0$ if $n\ge0,$ $R(\epsilon^n)=\epsilon^n$
if $n<0.$ If $\Gamma$ is primitive, as defined above, then $\phi(\Gamma)$
has only a simple pole in $\epsilon,$ hence $(1-R)\phi(\Gamma)$ is a good
renormalized value for $\Gamma.$ If $\Gamma$ does have subdivergences, the
situation is more complicated. However, the map $S_R^\phi: \mathcal{H}\rightarrow V$
\begin{equation*}
S_R^\phi(\Gamma) = -R\left(\phi(\Gamma)-\sum S_R^\phi(\Gamma')\phi(\Gamma'')\right)
\end{equation*}
provides the counterterm prescribed by the Bogoliubov recursion, and
$(S_R^\phi\ast\phi)(\Gamma)$ yields the renormalized value of $\Gamma.$
The map $S_R^\phi$ is a recursive deformation of $\phi\circ S$
by $R,$ compare its definition with (\ref{eq:antipode}). These are results
obtained by one of the authors in collaboration with Connes \cite{Kreimer,CK,CK2}.\\\\
For $S_R^\phi$ to be an algebra homomorphism again, one requires $R$ to
be a Rota-Baxter operator, studied in a more general setting by Ebrahimi-Fard, Guo and
one of the authors in \cite{EFIntegrable1,EFIntegrable2,EFSpitzer}.
The Rota-Baxter property is at the algebraic origin of the Birkhoff decomposition introduced
in \cite{CK2,CK3}. In the presence of mass terms, or gauge symmetries
etc. in the Lagrangian, $\phi,$ $S_R^\phi$ and $S_R^\phi\star\phi$ may contribute to
several form factors in the usual way. This can be resolved by considering a slight
extension of the Hopf algebra containing projections onto single structure functions,
as discussed for example in \cite{CK2,KreimerToronto}. For the case of gauge
theories, a precise definition of the coefficients $n(\gamma_1,\gamma_2,\Gamma)$
is given in \cite{Anatomy}.\\\\
The Hopf algebra $\mathcal{H}$ arises from the simple insertion of graphs
into each other in a completely canonical way. Indeed, the pre-Lie product
determines the coproduct, and the coproduct determines the antipode. Like this, each
quantum field theory gives rise to such a Hopf algebra $\mathcal{H}$ based on
its 1PI graphs. It is no surprise then that there is an even more universal
Hopf algebra behind all of them: The Hopf algebra $\mathcal{H}_{rt}$ of rooted trees
\cite{Kreimer,CK}. In order to see this, imagine a purely nested situation of
subdivergences like
\begin{equation*}
\phifull
\end{equation*}
which can be represented by the rooted tree
\begin{equation*}
\oc.
\end{equation*}
To account for each single graph of this kind, the tree's vertices should
actually be labeled according to which primitive graph they correspond to
(plus some gluing data) which we will suppress for the sake of simplicity.
The coproduct on $\mathcal{H}_{rt}$ -- corresponding to the one (\ref{eq:graphdelta})
of $\mathcal{H}$ -- is
\begin{equation*}
\Delta (\tau)=\mathbb{I}\otimes \tau+\tau\otimes\mathbb{I}+ \sum_{adm. c}
P_c(\tau)\otimes R_c(\tau)
\end{equation*}
where the sum runs over all \emph{admissible cuts} of the tree
$\tau.$ A cut of $\tau$ is a nonempty subset of its edges which are
to be removed. A cut $c(\tau)$ is defined to be admissible,
if for each leaf $l$ of $\tau$ at most one edge on the
path from $l$ to the root is cut. The product of subtrees which fall down when those edges
are removed is denoted $P_c(\tau).$ The part which remains connected with
the root is denoted $R_c(\tau).$ Here is an example:
\begin{eqnarray*}
\Delta\left(\ofork\right)&=&\ofork\otimes\mathbb{I}+\mathbb{I}\otimes\ofork+
2\bullet\otimes\olongline+\\
&+&\bullet\bullet\otimes\oline+\oc\otimes\bullet.
\end{eqnarray*}
Compared to $\mathcal{H}_{rt},$ the advantage of $\mathcal{H}$ is however
that overlapping divergences are resolved automatically. To achieve this
in $\mathcal{H}_{rt}$ requires some care \cite{KreimerOverlapping}.
\section{From Hochschild cohomology to physics}
There is a natural cohomology theory on $\mathcal{H}$ and $\mathcal{H}_{rt}$
whose non-exact 1-cocycles play an important ''operadic'' role in the sense
that they drive the recursion that define the full 1PI Green's functions in
terms of primitve graphs. In order to introduce this cohomology theory, let $A$
be any bialgebra. We view $A$ as a bicomodule over
itself with right coaction $(id\otimes\epsilon)\Delta.$ Then the Hochschild
cohomology of $A$ (with respect to the coalgebra part) is defined as follows
\cite{CK}:
Linear maps $L: A\rightarrow A^{\otimes n}$ are considered as
$n$-cochains. The operator $b,$ defined as
\begin{equation}\label{Hscb}
bL := (id\otimes L)\Delta+\sum_{i=1}^n(-1)^i\Delta_i
L+(-1)^{n+1}L\otimes \mathbb{I}
\end{equation}
furnishes a codifferential: $b^2 = 0.$ Here $\Delta$ denotes the
coproduct of $A$ and $\Delta_i$ the coproduct applied to the $i$-th
factor in $A^{\otimes n}$. The map
$L\otimes\mathbb{I}$ is given by $x\mapsto L(x)\otimes\mathbb{I}.$ Clearly
this codifferential encodes only information about the \emph{coalgebra} (as opposed to
the algebra) part of $A.$
The resulting cohomology is denoted $\operatorname{HH}^\bullet_\epsilon(A).$
For $n=1,$ the cocycle condition $bL=0$ is simply
\begin{equation}\label{eq:Hscb1}
\Delta L = (id\otimes L) \Delta+L\otimes\mathbb{I}
\end{equation}
for $L$ a linear endomorphism of $A.$ In the Hopf algebra $\mathcal{H}_{rt}$
of rooted trees (where things are often simpler), a 1-cocycle is quickly found:
the grafting operator $B_+,$ defined by
\begin{eqnarray*}
B_+(\mathbb{I})&=&\bullet\nonumber\\
B_+(\tau_1 \ldots\tau_n)&=&\ocf\quad\mbox{for trees }\tau_i
\end{eqnarray*}
joining all the roots of its argument to a newly created root. Clearly, $B_+$ reminds
of an operad multiplication. It is easily seen that $B_+$ is not exact and therefore
a generator (among others) of $\operatorname{HH}^1_\epsilon(\mathcal{H}_{rt}).$
Foissy \cite{FoissyBSM1} showed that $L\mapsto L(\mathbb{I})$ is an onto map
$\operatorname{HH}^1_\epsilon(\mathcal{H}_{rt})\rightarrow \operatorname{Prim}\mathcal{H}_{rt}.$
The higher Hochschild cohomology $(n\ge 2)$
of $\mathcal{H}_{rt}$ is known to vanish \cite{FoissyBSM1}. The pair
$(\mathcal{H}_{rt},B_+)$ is the universal model for all Hopf algebras of
Feynman graphs and their 1-cocycles \cite{CK}. Let us now turn to those 1-cocycles
of $\mathcal{H}.$ Clearly, every primitve graph $\gamma$ gives rise to a 1-cocycle
$B_+^\gamma$ defined as the operator which inserts its argument, a product of graphs, into
$\gamma$ in all possible ways. Here is a simple example:
\begin{equation*}
B_{+}^{\seone}\left(\vert\right) = \frac{1}{2}\setwo
\end{equation*}
See \cite{Anatomy} for the general definition involving some
combinatorics of insertion places and symmetries.\\\\
It is an important consequence of the $B_+^\gamma$ satisfying the cocycle condition (\ref{eq:Hscb1}) that
\begin{equation}\label{eq:locality}
(S_R^\phi\ast\phi)B_+ = (1-R)\tilde B_+(S_R^\phi\ast\phi)
\end{equation}
where $\tilde B_+$ is the push-forward of $B_+$ along the Feynman rules $\phi.$
In other words, $\tilde B_+^\gamma$ is the integral operator corresponding to the
skeleton graph $\gamma.$ This is the combinatorial key to the
proof of locality of counterterms and finiteness of renormalization
\cite{Collins,Kreimer2,BK,BK2}. Indeed, equation (\ref{eq:locality})
says that after treating all subdivergences, an overall
subtraction $(1-R)$ suffices. The only analytic ingredient is Weinberg's
theorem applied to the primitive graphs.
In \cite{BK} it is emphasized that $\mathcal{H}$ is actually generated
(and determined) by the action of prescribed 1-cocycles and the
multiplication. A version of (\ref{eq:locality}) with decorated trees
is available which describes renormalization in coordinate space
\cite{BK}. \\\\
The 1-cocycles $B_+^\gamma$ give rise to a number of useful Hopf subalgebras
of $\mathcal{H}.$ Many of them are isomorphic. They are studied in \cite{BK2}
on the model of decorated rooted trees, and we will come back to them in the
next section. In \cite{Anatomy} one of the authors showed that in nonabelian
gauge theories, the existence of a certain Hopf subalgebra, generated by
1-cocycles, is closely related to the Slavnov-Taylor identities for the couplings
to hold. In a similar spirit, van Suijlekom showed that, in QED,
Ward-Takahashi identities, and in nonabelian Yang-Mills theories, the Slavnov-Taylor
identities for the couplings generate Hopf ideals $\mathcal{I}$ of
$\mathcal{H}$ such that the quotients $\mathcal{H}/\mathcal{I}$ are defined and
the Feynman rules factor through them \cite{Suijlekom1,Suijlekom2}. The Hopf algebra
$\mathcal{H}$ for QED had been studied before in \cite{BKautom,KDusing,VP}.
\section{Dyson-Schwinger equations}
The ultimate application of the Hochschild 1-cocycles introduced in the previous
section aims at non-perturbative results. Dyson-Schwinger equations, reorganized
using the correspondence
$\operatorname{Prim}\mathcal{H}\rightarrow \operatorname{HH}_\epsilon^1(\mathcal{H}),$
become recursive equations in $\mathcal{H}[[\alpha]],$ $\alpha$ the coupling constant,
with contributions from (degree 1) 1-cocycles. The Feynman rules connect them to the
usual integral kernel representation. We remain in the massless $\phi^3$ theory in
6 dimensions for the moment. Let $\Gamma^\Yleft$ be the \emph{full} 1PI vertex function,
\begin{equation}\label{eq:OnePIvert}
\Gamma^\Yleft = \mathbb{I}+\sum_{\operatorname{res} \Gamma=\Yleft} \alpha^{|\Gamma|}\frac{\Gamma}
  {\operatorname{Sym}\Gamma}
\end{equation}
(normalized such that the tree level contribution equals 1). This is a formal power series in $\alpha$
with values in $\mathcal{H}.$ Here $\operatorname{res} \Gamma$
is the result of collapsing all internal lines of $\Gamma.$ The graph $\operatorname{res} \Gamma$
is called the residue of $\Gamma.$ In a renormalizable theory, $\operatorname{res}$ can be
seen as a map from the set of generators of $\mathcal{H}$ to the terms in the Lagrangian.
For instance, in the $\phi^3$ theory, vertex graphs have residue $\Yleft,$ and self energy
graphs have residue $-.$ The number $\operatorname{Sym}\Gamma$ denotes the order of the group
of automorphisms of $\Gamma,$ defined in detail for example in \cite{Anatomy,Suijlekom2}.
Similarly, the \emph{full} inverse propagator $\Gamma^-$ is represented by
\begin{equation}\label{eq:invprop}
\Gamma^- = \mathbb{I}-\sum_{\operatorname{res} \Gamma = -}\alpha^{|\Gamma|}\frac{\Gamma}{\operatorname{Sym}\Gamma}.
\end{equation}
These series can be reorganized by summing only over primitive graphs, with all possible
insertions into these primitive graphs. In $\mathcal{H},$ the insertions are afforded by
the corresponding Hochschild 1-cocycles. Indeed,
\begin{eqnarray}\label{eq:twodses}
\Gamma^{\Yleft} &=& \mathbb{I} +
\sum_{\gamma\in\operatorname{Prim}\mathcal{H},\operatorname{res}\gamma=\Yleft}
\frac{\alpha^{|\gamma|}B_+^\gamma(\Gamma^{\Yleft} Q^{|\gamma|})}{\operatorname{Sym}\gamma}\nonumber\\
\Gamma^{-} &=& \mathbb{I} -
\sum_{\gamma\in\operatorname{Prim}\mathcal{H},\operatorname{res}\gamma=-}
\frac{\alpha^{|\gamma|}B_+^\gamma(\Gamma^{-} Q^{|\gamma|})}{\operatorname{Sym}\gamma}.
\end{eqnarray}
The universal invariant charge $Q$ is a monomial in the $\Gamma^{r}$ and their
inverses, where $r$ are residues (terms in the Lagrangian) provided by the theory.
In $\phi^3$ theory we have $Q=(\Gamma^{\Yleft})^2(\Gamma^-)^{-3}.$ In
$\phi^3$ theory, the universality of $Q$ (i.~e.~ the fact that the same $Q$ is good
for \emph{all} Dyson-Schwinger equations of the theory) comes from a simple topological
argument.
In nonabelian gauge theories however, the universality of $Q$ takes care that
the solution of the corresponding system of coupled Dyson-Schwinger equations
gives rise to a Hopf subalgebra and therefore amounts to the Slavnov-Taylor
identities for the couplings \cite{Anatomy}.\\\\
The system (\ref{eq:twodses}) of coupled Dyson-Schwinger equations has
(\ref{eq:OnePIvert},\ref{eq:invprop}) as its solution. Note that in the first equation
of (\ref{eq:twodses}) an \emph{infinite} number of cocycles contributes as there
are infinitely many primitive vertex graphs in $\phi^3_6$ theory -- the second
equation has only finitely many contributions -- here one. Before we describe how to
actually attempt to solve equations of this kind analytically (application of the Feynman
rules $\phi$), we discuss the combinatorial ramifications of this construction in the
Hopf algebra. It makes sense to call all (systems of) recursive equations of the form
\begin{eqnarray*}
X_1 &=& \mathbb{I} \pm \sum_n \alpha^{k^1_n} B^{d^1_n}_+(M^1_n)\\
&\ldots&\\
X_s & = & \mathbb{I} \pm \sum_n \alpha^{k^s_n} B^{d^s_n}_+(M^s_n)
\end{eqnarray*}
\emph{combinatorial} Dyson-Schwinger equations, and to study their combinatorics.
Here, the $B^{d_n}_+$ are non-exact Hochschild 1-cocycles and the $M_n$ are
monomials in the $X_1\ldots X_s.$
In \cite{BK2} we studied a large class of single (uncoupled)
combinatorial Dyson-Schwinger equations in a decorated version of
$\mathcal{H}_{rt}$ as a model for vertex insertions:
\begin{equation*}
X=\mathbb{I}+\sum_{n=1}^\infty \alpha^n w_n  B_+^{d_n}(X^{n+1})
\end{equation*}
where the $w_n\in\mathbb{Q}.$ For example, $X=\mathbb{I}+\alpha B_+(X^2)+\alpha^2 B_+(X^3)$
is in this class. It turns out \cite{Kreimer2,BK2} that the coefficients $c_n$ of $X,$
defined by $X = \sum_{n=0}^\infty \alpha^n c_n$ generate a Hopf subalgebra themselves:
\begin{equation*}
\Delta(c_n)= \sum_{k=0}^n P^n_k\otimes c_k.
\end{equation*}
The $P^n_k$ are homogeneous polynomials of degree $n-k$ in the
$c_l,$ $l\le n.$ These polynomials have been worked out explicitly in \cite{BK2}.
One notices in particular that the $P^n_k$ are independent of the
$w_n$ and $B_+^{d_n},$ and hence that under mild assumptions (on the
algebraic independence of the $c_n$) the Hopf subalgebras
generated this way are actually isomorphic. For example,
$X=\mathbb{I}+\alpha B_+(X^2)+\alpha^2 B_+(X^3)$ and $X=\mathbb{I}+\alpha B_+(X^2)$
yield isomorphic Hopf subalgebras. This is an aspect of the fact that
\emph{truncation} of Dyson-Schwinger equations -- considering only a finite
instead of an infinite number of contributing cocycles -- does make
(at least combinatorial) sense. Indeed, the combinatorics remain invariant.
Similar results hold for Dyson-Schwinger equations in the true Hopf algebra
of graphs $\mathcal{H}$ where things are a bit more difficult though as the
cocycles there involve some bookkeeping of insertion places.\\\\
The simplest nontrivial Dyson-Schwinger equation one can think of is the
\emph{linear} one:
\begin{equation*}
X = \mathbb{I}+\alpha B_+(X).
\end{equation*}
Its solution is given by $X=\sum_{n=0}^\infty\alpha^n (B_+)^n (\mathbb{I}).$
In this case $X$ is grouplike and the corresponding Hopf subalgebra of $c_n$s is
\emph{cocommutative} \cite{KreimerEtude}. A typical and important \emph{non-linear}
Dyson-Schwinger equation arises from propagator insertions:
\begin{equation*}
X=\mathbb{I}-\alpha B_+(1/X),
\end{equation*}
for example the massless fermion propagator in Yukawa theory where only the
fermion line obtains radiative corrections (other corrections are ignored).
This problem has been studied and solved by Broadhurst and one of the authors
in \cite{BroadhurstKreimerDSE} and revisited recently by one of the authors
and Yeats \cite{KY}. As we now turn to the analytic aspects of Dyson-Schwinger
equations, we briefly sketch the general approach presented in \cite{KY} on how to
successfully treat the nonlinearity of Dyson-Schwinger equations. Indeed,
the \emph{linear} Dyson-Schwinger equations can be solved by a simple scaling ansatz
\cite{KreimerEtude}. In any case, let $\gamma$ be a primitive graph. The following
works for amplitudes which depend on a single scale, so let us assume a massless
situation with only one non-zero external momentum -- how more than one external
momentum (vertex insertions) are incorporated by enlarging the set of primitive
elements is sketched in \cite{KreimerToronto}. The grafting operator $B_+^\gamma$
associated to $\gamma$ translates to an integral operator under the (renormalized)
Feynman rules
\begin{equation*}
\phi_R(B_+^\gamma)(\mathbb{I})(p^2/\mu^2) = \int (I_\gamma(k,p)-I_\gamma(k,\mu)) dk
\end{equation*}
where $I_\gamma$ is the integral kernel corresponding to $\gamma,$ the internal
momenta are denoted by $k,$ the external momentum by $p,$ and $\mu$ is the fixed
momentum at which we subtract: $R(x)=x|_{p^2=\mu^2}.$ \\\\
In the following we stick to the special case discussed in \cite{KY} where only
\emph{one} internal edge is allowed to receive corrections.
The integral kernel $\phi(B_+^\gamma)$ defines a Mellin transform
\begin{equation*}
F(\rho) = \int I_\gamma(k,\mu) (k_i^2)^{-\rho} dk
\end{equation*}
where $k_i$ is the momentum of the internal edge of $\gamma$ at which insertions
may take place (here the fermion line). If there are several insertion sites,
obvious multiple Mellin transforms become necessary. The case of two (propagator)
insertion places has been studied, at the same example, in \cite{KY}. \\\\
The function $F(\rho)$ has a simple pole in $\rho$ at 0. We write
\begin{equation*}
F(\rho) = \frac{r}{\rho}+\sum_{n=0}^\infty f_n \rho^n
\end{equation*}
We denote $L= \log p^2/\mu^2.$ Clearly $\phi_R(X)=1+\sum_n \gamma_n L^n.$
An important result of \cite{KY} is that, even in the difficult nonlinear
situation, the anomalous dimension $\gamma_1$ is implicitly defined by the
residue $r$ and Taylor coefficients $f_n$ of the Mellin transform $F.$ On the
other hand, all the $\gamma_n$ for $n\ge 2,$ are recursively defined in terms
of the $\gamma_i,$ $i<n.$ This last statement amounts to a renormalization
group argument that is afforded in the Hopf algebra by the scattering formula
of \cite{CK3}. Curiously, for this argument only a linearized part of the coproduct
is needed. We refer to \cite{KY} for the actual algorithm.
For a \emph{linear} Dyson-Schwinger equation, the situation is considerably
simpler as the $\gamma_n=0$ for $n\ge 2$ since $X$ is grouplike \cite{KreimerEtude}.
\\\\
Let us restate the results for the high energy sector of non-linear Dyson-Schwinger
equations \cite{BroadhurstKreimerDSE,KY}: Primitive graphs $\gamma$ define Mellin
transforms via their integral kernels $\tilde B^\gamma_+.$ The anomalous dimension
$\gamma_1$ is \emph{implicitly} determined order by order from the coefficients of
those Mellin transforms. All non-leading log coefficients $\gamma_n$ are recursively
determined by $\gamma_1,$ thanks to the renormalization group. This reduces, in
principle, the problem to a study of all the primitive graphs and the intricacies of
insertion places. \\\\
Finding useful representations of those Mellin transforms -- even
one-dimensional ones -- of higher loop order skeleton graphs is difficult.
However, the two-loop primitive vertex in massless Yukawa theory
has been worked out by Bierenbaum, Weinzierl and one of the authors in \cite{BKW}, a result that can be applied to
other theories as well. Combined with the algebraic treatment
\cite{BroadhurstKreimerDSE,BK2,KY} sketched in the previous paragraphs
and new geometric insight on primitive graphs (see section \ref{s:periods}),
there is reasonable hope that actual solutions of Dyson-Schwinger equations
will be more accessible in the future. \\\\
Using the Dyson-Schwinger analysis, one of the authors and Yeats \cite{KY2}
were able to deduce a bound for the convergence of superficially divergent
amplitudes/structure functions from the (desirable) existence of a bound
for the superficially convergent amplitudes. \\\\
\section{Feynman integrals and periods of mixed (Tate) Hodge
structures} \label{s:periods}
A primitive graph $\Gamma\in \operatorname{Prim} \mathcal{H}$
defines a real number $r_\Gamma$, called the \emph{residue} of
$\Gamma$, which is independent of the renormalization scheme. In the
case that $\Gamma$ is massless and has one external momentum $p,$
the residue $r_\Gamma$ is the coefficient of $\log p^2/\mu^2$ in
$\phi_R(\Gamma)=(1-R)\phi(\Gamma).$ It coincides with the
coefficient $r$ of the Mellin transform introduced in the previous
section. One may ask what kind of a number $r$ is, for example if it
is rational or algebraic. The origin of this question is that the
irrational or transcendental numbers that show up for various
$\Gamma$ strongly suggest a motivic interpretation of the $r_\Gamma.$
Indeed, explicit calculations \cite{BroadKreimer3,Broadhurst-Kreimer-96-2,BGK}
display patterns of Riemann zeta and multiple zeta values that are
known to be periods of mixed Tate Hodge structures -- here the
periods are provided by the Feynman rules which produce
$\Gamma\mapsto r_\Gamma.$ By disproving a related conjecture of
Kontsevich, Belkale and Brosnan \cite{BB} have shown that not all
these Feynman motives must be mixed Tate, so one may expect a larger
class of Feynman periods than multiple zeta values.
Our detailed understanding of these phenomena is still far from
complete, and only some very first steps have been made in the last
few years. However, techniques developed in recent work by Bloch,
Esnault and one of the authors \cite{BEK} do permit reasonable
insight for some special cases which we briefly sketch in the following.\\\\
Let $\Gamma$ be a logarithmically divergent massless primitive graph
with one external momentum $p$. It is convenient to work in the
''Schwinger'' parametric representation \cite{IZ} obtained by the
usual trick of replacing propagators
\begin{equation*}
\frac{1}{k^2} = \int_0^\infty d a e^{-ak^2},
\end{equation*}
and performing the loop integrations (Gaussian integrals) first
which leaves us with a (divergent) integral over various Schwinger
parameters $a.$ It is a classical exercise \cite{IZ,BEK,BlochJapan}
to show that in four dimensions, up to some powers of $i$ and
$2\pi,$
\begin{equation*}
\phi(\Gamma) = \int_0^\infty da_1\ldots da_n
\frac{e^{-Q_\Gamma(a,p^2)/\Psi_\Gamma(a)}}{\Psi_\Gamma^2(a)}
\end{equation*}
where $n$ is the number of edges of $\Gamma.$ $Q_\Gamma$ and
$\Psi_\Gamma$ are graph polynomials of $\Gamma$, where
$\Psi_\Gamma,$ sometimes called \emph{Symanzik} or \emph{Kirchhoff
polynomial}, is defined as follows: Let $T(\Gamma)$ be the set of
spanning trees of $\Gamma,$ i.~e.~ the set of connected simply
connected subgraphs which meet all vertices of $\Gamma.$ We think of
the edges $e$ of $\Gamma$ as being numbered from 1 to $n.$ Then
\begin{equation*}
\Psi_\Gamma = \sum_{t\in T(\Gamma)} \prod_{e\not\in t} a_e
\end{equation*}
This is a homogeneous polynomial in the $a_i$ of degree
$|H_1(\Gamma)|.$ It is easily seen (scaling behaviour of $Q_\Gamma$
and $\Psi_\Gamma)$ that
$r_\Gamma=\frac{\partial\phi_R(\Gamma)}{\partial \log p^2/\mu^2}$ is
extracted from $\phi(\Gamma)$ by considering the $a_i$ as
homogeneous coordinates of $\mathbb{P}^{n-1}(\mathbb{R})$ and evaluating at
$p^2=0:$
\begin{equation}\label{eq:period}
r_\Gamma = \int_{\sigma\subset \mathbb{P}^{n-1}(\mathbb{R})}
\frac{\Omega}{\Psi_\Gamma^2}
\end{equation}
where $\sigma=\{[a_1,\ldots,a_n]: \mbox{ all } a_i \mbox{ can be
choosen }\ge 0\}$ and $\Omega$ is a volume form on
$\mathbb{P}^{n-1}.$ Let $X_\Gamma:=\{\Psi_\Gamma =
0\}\subset\mathbb{P}^{n-1}.$ If $|H_1(\Gamma)|=1,$ the integrand in
(\ref{eq:period}) has no poles. If $|H_1(\Gamma)|>1,$ poles will
show up on the union $\Delta=\bigcup_{\gamma\subsetneq \Gamma,
H_1(\gamma) \neq 0} L_\gamma$ of coordinate linear spaces
$L_\gamma=\{a_e = 0$ for $e$ edge of $\gamma\}$
 -- these need to be separated from the chain of
integration by blowing up. The blowups being understood, the Feynman
motive is, by abuse of notation,
\begin{equation*}
H^{n-1}(\mathbb{P}^{n-1}-X_\Gamma,\Delta-\Delta\cap X_\Gamma)
\end{equation*}
with Feynman period given by (\ref{eq:period}). See
\cite{BEK,BlochJapan} for details. Some particularly accessible
examples are the \emph{wheel with $n$ spokes graphs}
\begin{equation*}
\Gamma_n := \wheel
\end{equation*}
studied extensively in \cite{BEK}. The corresponding Feynman periods
(\ref{eq:period}) yield rational multiples of zeta values
\cite{BroadKreimer3}
\begin{equation*}
r_{\Gamma_n} \in \zeta(2n-3)\mathbb{Q}.
\end{equation*}
Due to the simple topology of the $\Gamma_n,$ the geometry of the
pairs $(X_{\Gamma_n},\Delta_{\Gamma_n})$ are well understood and the
corresponding motives have been worked out explicitly \cite{BEK}.
The methods used are however nontrivial and
not immediately applicable to more general situations. \\\\
When confronted with non-primitive graphs, i.~e.~ graphs with
subdivergences, there are more than one period to consider. In the
Schwinger parameter picture, subdivergences arise when poles appear
along exceptional divisors as pieces of $\Delta$ are blown up. This
situation can be understood using limiting mixed Hodge structures
\cite{BlochJapan}, see also \cite{KreimerNumbers,MK2} for a toy
model approach to the combinatorics involved. In \cite{BlochJapan}
it is also shown how the Hopf algebra $\mathcal{H}$ of graphs lifts
to the category of motives. For the motivic role of solutions of
Dyson-Schwinger equations we refer to work in progress. Finally we
mention that there is related work by Connes and Marcolli
\cite{CoMa2,CoMa3} who attack the problem via Riemann-Hilbert
correspondences and motivic Galois theory.\\\\
\bf Acknowledgements. \rm We thank Spencer Bloch and Karen Yeats for
discussion on the subject of this review. The first named author (C.~B.)
thanks the organizers of the ICMP 2006 and the IHES for general support. His
research is supported by the Deutsche Forschungsgemeinschaft. The
IHES, Boston University and the Erwin-Schr\"odinger-Institute are
gratefully acknowledged for their kind hospitality. At the time
of writing this article, C.~B.~ is visiting the ESI as a Junior
Research Fellow.

%
%
\bibliographystyle{plain}
\bibliography{lit}

\end{document}